\documentclass[12 pt]{article}

\usepackage{amsmath,amstext,amsgen,amsbsy,amsopn,amsfonts,graphicx,amssymb,
rotating}

\begin{document}

\title{Bifurcations in One Degree-of-Vibration Quantum Billiards}

\author{Mason A. Porter and Richard L. Liboff \\ \\ Center for Applied 
Mathematics \\ and \\ Schools of Electrical Engineering and Applied Physics \\ 
\\ Cornell University}

\date{April, 2000}

\maketitle

\begin{center}
\section*{Abstract}
\end{center}

\vspace{.08 in}

	We classify the local bifurcations of one 
\begin{itshape}dov\end{itshape} quantum billiards, showing that only 
saddle-center bifurcations can occur.  We analyze the resulting planar system 
when there is no coupling in the superposition state.   In so doing, we also 
consider the global bifurcation structure.  Using a double-well potential as a
 representative example, we demonstrate how to locate bifurcations in 
parameter space.  We also discuss how to approximate the cuspidal loop using 
AUTO as well as how to cross it via continuation by detuning the dynamical 
system. Moreover, we show that when there is coupling, the resulting 
five-dimensional system---though chaotic---has a similar underlying structure.
  We verify numerically that both homoclinic orbits and cusps occur and 
provide an outline of an analytical argument for the existence of such 
homoclinic orbits.  Small perturbations of the system reveal homoclinic 
tangles that typify chaotic behavior.	

\vspace{.25 in}

\subsection*{MSC NOS 37N20, 37K55, 37M20}

\vspace{.25 in}

\section{Introduction}

\vspace{.1 in}

	Quantum chaos is an underdeveloped area of dynamical systems theory.  
One purpose of studying it is to generalize the notions of classical 
Hamiltonian chaos to the quantum regime, which has not yet been done in a 
universally accepted manner.  One type of quantum chaos is often called 
semiquantum chaos, as these systems consist of classical (Hamiltonian) 
variables coupled with quantum variables.\cite{atomic}  Vibrating quantum 
billiards are a representative example of semiquantum chaotic 
systems.\cite{sazim,nec}  They may be used as models for quantum well 
microdevice components (such as quantum dots and quantum wires), Fermi 
accelerators\cite{fermi}, and intranuclear particle behavior.

	In the present paper, we consider the bifurcation structure of 
vibrating billiard systems.  Quantum billiards describe the motion of a point 
particle undergoing perfectly elastic collisions in a bounded domain.  The 
particle's motion is described by the Schr\"odinger equation with Dirichlet 
boundary conditions.  One defines the ``degree-of-vibration'' 
(\begin{itshape}dov\end{itshape}) of a billiard as the number of boundary 
dimensions that vary with time.  If the boundary is time-independent, the 
billiard is said to have zero \begin{itshape}dov\end{itshape}.  The linear 
vibrating billiard and the radially vibrating spherical billiard have a single
 \begin{itshape}dov\end{itshape}, and the rectanglular billiard with 
time-varying length and width has two \begin{itshape}dov\end{itshape}.  

	A zero \begin{itshape}dov\end{itshape} quantum billiard exhibits only 
integrable behavior if it is globally separable.\cite{gutz}  It must, for 
example, be convex and describable by a non-composite 
geometry.\cite{katok,mac}  If part of the billiard is concave (or is 
composite, like the stadium billiard), it may behave chaotically, as it shares
 many of the instability properties of the Anosov diffeomorphism.  Bl\"umel 
and Esser\cite{vibline} found quantum chaos in the one-dimensional vibrating 
quantum billiard.  Liboff and Porter\cite{sazim} extended these results 
to spherical quantum billiards with vibrating surfaces and derived necessary 
conditions for chaotic behavior.  They also generalized their results to other
 one \begin{itshape}dov\end{itshape} billiards.\cite{nec}  The purpose of the 
present paper is to examine bifurcations in single 
\begin{itshape}dov\end{itshape} quantum billiards that occur when one alters 
the potential in which the billiard resides.
	
	Vibrating quantum billiards are important for several reasons.  Though
 an idealized model, they are nevertheless useful for the study of quantum 
chaos.  From a more practical standpoint, vibrating quantum billiards may be 
applied to several problems in physics.  The radially vibrating spherical 
quantum billiard, for example, may be used as a model for particle behavior in
 the nucleus\cite{wong} as well as for the quantum dot microdevice 
component.\cite{qdot}.  Additionally, the vibrating cylindrical billiard may 
be used as a model for the quantum wire, another microdevice 
component.\cite{qwire}  Other geometries of vibrating quantum billiards have 
similar applications.  They may also be used as models of Fermi 
acceleration of cosmic rays.\cite{fermi}  The study of quantum chaos in 
vibrating billiard systems is thus important both because it expands the 
mathematical theory of dynamical systems and because it can be applied to 
problems in nuclear and mesoscopic physics.

	In the present paper, we show that saddle-centers are the only type of
 bifurcations that can occur in one \begin{itshape}dov\end{itshape} quantum 
billiards.  When there is no coupling in the superposition state, we show how 
to analyze the resulting planar system analytically and numerically.  
Considering a double-well potential as a representative example, we 
demonstrate how to locate bifurcations in parameter space.  We also discuss 
how to approximate the cuspidal loop using AUTO as well as how to continue 
past it by detuning the dynamical system.  We also mention a shooting method 
for a more detailed analysis of the cuspidal loop.  Moreover, we show that 
when there is coupling, the resulting five-dimensional system---though 
chaotic---has a similar underlying structure.  We verify numerically that both
 homoclinic orbits and cusps occur, and we outline an analytic argument that 
demonstrates the existence of homoclinic orbits.  Small perturbations of the 
system reveal homoclinic tangles that typify chaotic behavior.

\vspace{.1 in}

\section{Equations of Motion}
	
	The goal of the present project is to examine the behavior of one 
\begin{itshape}dov\end{itshape} quantum billiards in various potentials in 
order to determine the effects of the potential on the behavior of the system.
  In particular, we analyze bifurcations of equilibria both analytically and 
numerically.  We consider a two-state superposition solution to the vibrating 
billiard, and we examine the above problem for both the case in which the two 
states experience coupling and that in which they do not.

	The present problem is described by the Schr\"odinger equation with 
solutions that are constrained to vanish on a time-dependent boundary 
$a(t)$.\cite{nec}  That is,
\begin{equation}
	\psi(r,t;a(t)) = 0 \text{       for       } r = a(t).
\end{equation}
	Because of the time-dependence of the boundary, the above boundary 
condition is nonlinear.  The (mathematical) problem at hand is to find a 
boundary $a(t)$ such that Dirichlet boundary conditions are satisfied on it.  
One can then, in principle, insert $a(t)$ into the eigenfunctions in our normal
 mode expansion of the wave $\psi(r,t;a(t))$ in order to obtain nonlinear 
normal modes.  When the radius $a(t)$ behaves chaotically, the nonlinear 
normal modes are examples of quantum-mechanical wave chaos.  One derives 
coupled nonlinear ordinary differential equations for $a(t)$ (and other 
variables) using a Gal\"erkin approximation.\cite{gucken,infinite}  
Considering a two-term superposition state then corresponds to taking a 
two-term Gal\"erkin projection.  The equations of motion that one obtains 
depend on whether a particular subset of the quantum numbers in the two states
 are the same.\cite{nec}  For the case of the radially vibrating sphere, the 
quantum numbers in question are the orbital and azimuthal quantum numbers $l$ 
and $m$, respectively.\cite{sazim}

	If these quantum numbers are the same in the two states, there is a 
coupling between them.  The evolution of the system is then described by
\begin{equation}
	\dot{x} = -\frac{\omega_0 y}{a^2} - \frac{2 \mu P z}{Ma}, 
\label{xdot}
\end{equation}
\begin{equation}
	\dot{y} = \frac{\omega_0 x}{a^2},
\end{equation}
\begin{equation}
	\dot{z} = \frac{2 \mu P x}{Ma},
\end{equation}
\begin{equation}
	\dot{a} = \frac{P}{M}, 
\end{equation}
and
\begin{equation}
	\dot{P} = -\frac{\partial V}{\partial a} + \frac{2[\epsilon_+ + 
\epsilon_-(z - \mu x)]}{a^3}, \label{Pdot}
\end{equation}
where $x$, $y$, and $z$ are Bloch variables\cite{bloch}
\begin{gather}
	x = \rho_{12} + \rho_{21}, y = i(\rho_{21} - \rho_{12}), z = \rho_{22}
 - \rho_{11} \label{blo}, \notag \\ x^2 + y^2 + z^2 = 1,
\end{gather}
$\rho_{mn} \equiv A_m A_n^*$ is the density matrix\cite{liboff}, $a$ is a 
displacement, $P$ is its conjugate momentum, $M$ is the mass of the billiard, 
$m \ll M$ is the mass of the confined particle, $\mu > 0$ is the coupling 
coefficient between the two eigenstates, $V = V(a)$ is the potential of the 
billiard boundary, $\omega_0 \equiv \frac{\epsilon_2 - \epsilon_1}{\hbar}$, 
$\epsilon_{\pm} \equiv \frac{(\epsilon_2 \pm \epsilon_1)}{2}$, and 
$\epsilon_1$ and $\epsilon_2$ $(\epsilon_2 \geq \epsilon_1)$ are the energies 
of the two states.  It has been shown that these equations exhibit quantum 
chaotic behavior.\cite{nec}

	If there is no coupling between the two eigenstates, the evolution of 
the system is described by a one degree-of-freedom Hamiltonian.  The equations
 of motion are
\begin{gather}
	\dot{a} = \frac{P}{M}, \notag \\  \dot{P} = 
-\frac{\partial V}{\partial a} + \frac{\lambda}{a^3}, \label{2dim}
\end{gather}
where 
\begin{equation}
	\lambda \equiv 2\left( \epsilon_1 |C_1|^2 + \epsilon_1 |C_2|^2 
\right), \label{lam}
\end{equation}
$C_1$ and $C_2$ are constants such that $|C_1|^2 + |C_2|^2 = 1$.  
The energy parameter $\lambda$ is necessarily positive because $\epsilon_i > 
0$ and the $|C_i|^2$ correspond to probabilities.

\section{Integrable Case: Absence of Coupling}

\vspace{.05 in}

	For the planar case, all $(a,P)$ that satisfy $\dot{a} = \dot{P} = 0$ 
are equilibrium points.  Each one is of the form $(\bar{a},0)$, where 
$\bar{a}$ satisfies
\begin{equation}
	\frac{\partial V}{\partial a}\left(\bar{a},0\right) = 
\frac{\lambda}{{\bar{a}}^3}.
\end{equation}
The eigenvalues of the integrable system (\ref{2dim}, \ref{lam}) at the 
stationary point $(\bar{a},0)$ are
\begin{equation}
	\sigma = \pm \sqrt{- \frac{1}{M} \left( \frac{\partial^2 V}
{\partial a^2}\left(\bar{a},0 \right) + \frac{3 \lambda}{\bar{a}^4}\right)}.
\end{equation}
These eigenvalues are either real with opposite sign or are pure imaginary 
pairs, so in the linear analysis, each equilibrium is either a center or a 
saddle point.  If 
\begin{equation}
	A \equiv \frac{\partial^2 V}{\partial a^2}\left(\bar{a},0 \right) + 
\frac{3 \lambda}{\bar{a}^4} > 0,
\end{equation}
then every equilibrium point is a linear center.  This holds, in particular, 
when the potential $V(a)$ has a single minimum (single-well potentials).  
Previous studies have focused on the harmonic potential
\begin{equation}
	V(a) = \frac{\mathcal{V}_2}{a_0^2}(a - a_0)^2 \equiv V_2(a-a_0)^2.
\end{equation}
Another interesting single-well potential is the quartic potential
\begin{equation}
	V(a) = \frac{\mathcal{V}_4}{a_0^4}(a - a_0)^4 \equiv V_4(a-a_0)^4. 
\label{quart}
\end{equation}
In the above equations, the $\mathcal{V}_i$ are dimensionless parameters.

	It is insightful to examine the evolution (particularly in the chaotic
 case) of the quantum billiard system with the above two potentials and derive
 a mechanical anology in terms of spring stiffness.  Given the same initial 
conditions and the quartic and quadratic potentials above (and assuming $V_2 =
 V_4$ for ease for comparison), one observes that the phase-plane trajectory 
described by the evolution of $a$ and $P$ for the quartic potential 
(\ref{quart}) has a larger radius of curvature (that is, a smaller curvature).
  For all initial conditions, the trajectory in the quadratic potential has a 
larger maximum $a$.  For initial conditions with sufficiently small $a(0)$, 
the quadratic potential induces trajectories with smaller maximum $|P|$, but 
the quartic potential eventually gives a larger maximum $|P|$ as $a(0)$ is 
increased.  

	An equilibrium for which $A < 0$ is a saddle point.  Since the present
 system is a Hamiltonian system, it is invariant under reflection about the 
$a$-axis, so that the eigenvectors representing the local stable and unstable 
manifolds are mapped to each other under this reflection.  Since the 
only other possible types of equilibria are centers, it follows that if there 
is at least one saddle point, the system must have saddle connections.  If 
there is exactly one, the connection is a homoclinic orbit, and there must be 
two of them emanating from the saddle because the system must have a center 
(in the right-half plane) on each side of the saddle.

	As one considers increasingly excited states of the system 
(corresponding to larger quantum numbers), the energy parameter $\lambda$ is 
increased.  Each saddle will eventually become a center as this occurs.  The 
quantity $A$ vanishes at such an equilibrium point.  The stationary point then
 has a double zero eigenvalue with the Jacobian
\begin{equation}  
	\begin{pmatrix}
		0 & 1 \\ 0 & 0
	\end{pmatrix}
\end{equation}
so that the saddle-center bifurcation that occurs has codimension two and 
gives rise to a global bifurcation corresponding to the breaking of the 
separatrix.\cite{hale,gucken}  To find the conditions satisfied at this point 
one can either solve the simultaneous equations $\dot{a} = 0, \sigma = 0$ or 
find the points at which the Hamiltonian has a double zero (which is 
equivalent to solving the system of equations $H(\bar{a},0) = 0, 
\frac{\partial H}{\partial a}\left(\bar{a},0 \right) = 0$).   If the potential 
has a constant term $V_0$, it does not change the evolution of the system 
since it does not appear in the equations of motion.  We thus let $V_0 = 0$ 
without loss of generality.  One finds that a saddle-center bifurcation occurs
 when
\begin{equation}
	\bar{\lambda} = \bar{a}^3\frac{\partial V}{\partial a} 
\left(\bar{a},0 \right)
\end{equation}
at the point $(\bar{a},0)$ satisfying
\begin{equation}
	\frac{\partial V}{\partial a} \left(\bar{a},0 \right) = 
-\frac{\bar{a}}{3} \frac{\partial ^2 V}{\partial a^2} \left(\bar{a},0 \right).
 \label{sad}
\end{equation}
Any solution to (\ref{sad}) giving $\lambda < 0$ is discarded as nonsensical.

	At the saddle-center bifurcation point, the stable and unstable 
eigenvectors of the equilibrium coincide along the $a$-axis, so that the 
stable and unstable manifolds overlap near the stationary point.  This cusp 
causes difficulties in numerical continuation attempts, as standard 
continuation techniques fail for this bifurcation.  One observes that the two 
homoclinic orbits that exist when $A < 0$ have coalesced into one.  (As $A$ 
increases, the homoclinic orbit on the left shrinks, becoming a single point 
at the saddle-center.  The orbit has infinite derivative with respect to 
arclength at the saddle-center point.)

\begin{figure}[htb] 
	\begin{centering}
		\leavevmode
		\includegraphics[width = 2.5 in, height = 3 in]{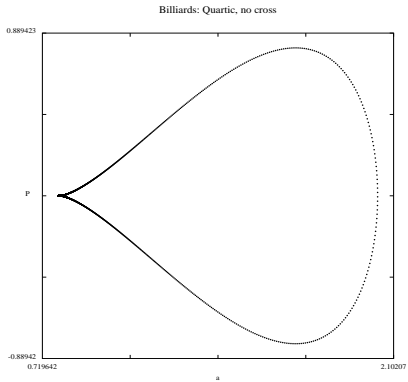}

\vspace{-.5 in}

		\caption{Approximate cuspidal homoclinic orbit.  The initial 
point used was $(0.7886751,0.001)$, just above the equilibrium.} \label{cusp}
	\end{centering}
\end{figure}

	As a specific example of this phenomenon, we consider the quartic 
potential
\begin{equation}
	V(a) = V_4(a-a_0)^4 + V_3(a-a_0)^3 + V_2(a-a_0)^2 + V_1(a-a_0),
\end{equation}
where $V_i \equiv \frac{\mathcal{V}_i}{a_0^i}$.  This potential has either one 
well or two.  In the latter case, there is a single saddle-center bifurcation 
point.  One can find $\bar{a}$ and $\bar{\lambda}$ exactly in this case, since
 the former is a root for a degree-three polynomial.  We initially restrict 
ourselves to the case in which $V_3 = V_1 = 0$, since all the dynamics of 
interest remain in this simpler case.  Note that the potential is symmetric 
about $a_0$.  For numerical purposes, consider the special case corresponding 
to the parameter values $a_0 = 1, V_2 = -1,$ and $V_4 = 1$.  
There is a saddle-center bifurcation at $\bar{\lambda} = \frac{1}{972}\left[3 
+ \sqrt{3}\right]^3 \sqrt{3} \approx 0.1888176.$  The corresponding stationary
 point is $\left(\frac{1}{2} + \frac{\sqrt{3}}{6},0 \right) \approx 
(0.7886751,0)$.  Using DsTool\cite{dstool}, we plotted an approximation of the
 homoclinic orbit emanating from this equilibrium (Fig. \ref{cusp}).

	More generally, one can consider even polynomials of higher degree in 
order to examine vibrating quantum billiards in an $N$-well potential.  If the 
polynomial is of degree six or higher, one may not be able to solve for 
$\bar{a}$ exactly in terms of radicals by Galois theory\cite{dummit}, since 
equation (\ref{sad}) is polynomial of degree at least five.  If its degree is 
exactly five, one can solve for $\bar{a}$ exactly in terms of elliptic 
functions.\cite{elliptic}

	Consider the problem of starting at $\lambda < \bar{\lambda}$ and 
attempting to continue along the bifurcation curve past the saddle-center.  
  Using AUTO\cite{auto,orbit} and the homotopy method encoded in HomCont, we 
followed the two homoclinic orbits for $\lambda = 0.15$ (Fig. \ref{homo}).  
The saddle connection in the present system has a codimension greater than 
one, as both regularity and non-degeneracy conditions are both 
violated.\cite{homcont,detect}  The present situation is 
\begin{itshape}degenerate\end{itshape} because for all $\lambda < 
\bar{\lambda}$, there are two homoclinic orbits emanating from the saddle 
point.  \begin{itshape}Regularity\end{itshape} is violated because the saddle 
point's two eigenvalues are negatives of each other.  (Moreover, their 
eigenvectors are related by reflection across the $a$-axis, since the system is
 Hamiltonian.) 

\begin{figure}[htb] 
	\begin{centering}
		\leavevmode
		\includegraphics[width = 2.5 in, height = 2 in]{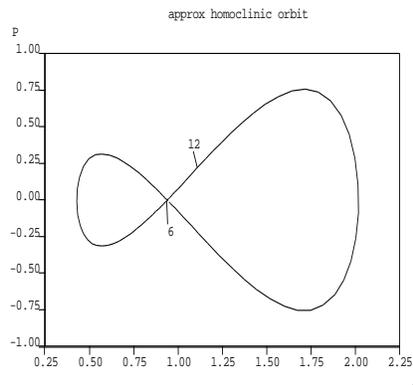}

\vspace{-.25 in}

		\caption{Homoclinic orbits emanating from $(0.8916637,0)$ for 
$\lambda = 0.15$.  The label 12 refers to the right homoclinic orbit, and the 
label 6 refers to the left one.} \label{homo}
	\end{centering}
\end{figure}

	Because the present system is degenerate and irregular, one cannot 
continue (in $\lambda$) past the saddle-center directly, as described in the 
AUTO manual.\cite{auto}  Hamiltonian systems possess a continua of homoclinic 
orbits, and there are numerical schemes that allow one to handle this 
phenomenon.  One can exploit the reversibility of the system by computing only
 half of a given saddle connection.  However, the cusp at the saddle-center 
point prevents this from working for the present system.  AUTO is incapable of
 continuing past a cusp, because $\frac{\partial P}{\partial a}$ vanishes 
there and the pseudo-arclength step must become vanishingly small for such a 
continuation step to be successful.  Because of machine precision, this cannot
 occur, and so one must ``detune'' the system to continue across the cusp.

	In general, Hamiltonian systems are described by 
\begin{equation}
	\dot{x} = J \nabla H(x,\lambda), \text{       }x \in \mathbb{R}^{2n},
\end{equation}
where
\begin{equation}
	J \equiv \begin{pmatrix}
			0 & I \\ -I & 0
		 \end{pmatrix}
\end{equation}
is the canonical $2n \times 2n$ symplectic matrix\cite{ms}.  One can detune 
the system by using a small perturbation parameter $\epsilon$ as 
follows.\cite{homcont}  The perturbed dynamical system,
\begin{equation}
	\dot{x} = J \nabla H(x,\lambda) + \epsilon \nabla H(x,\lambda),
\end{equation}
is no longer Hamiltonian, but the perturbation was constructed so that the 
locations of all equilibrium points are preserved.  With this detuning, the 
saddle-center bifurcation becomes a saddle-node bifurcation (the eigenvalues 
of the stationary point are now of the form $a \pm \sqrt{b'} \text{  } 
(a \neq 0)$ rather than of the form $\pm \sqrt{b}$).  One can then continue 
$\lambda$ past this point using AUTO (without utilizing the HomCont package). 
 Using this technique, one can compute the value of $\lambda$ at which the 
saddle-center bifurcation occurs as well as the cusp point of the homoclinic 
orbit corresponding to that value.  Moreover, once one has successfully 
continued past the cusp, one can simply let $\epsilon \longrightarrow 0$ and 
thereby work with the system when $\lambda > \bar{\lambda}$.  This method of 
continuation is useful as long as one needs to get past the cusp rather than 
do computations at the cusp itself.  For the present system, we used $\epsilon
 < 0$, since in that case the equilibria that are centers for $\epsilon = 0$ 
become stable spirals.  The continuation curve (in $\lambda$) is shown in 
Figure \ref{continue}.

\begin{figure}[htb] 
	\begin{centering}
		\leavevmode
		\includegraphics[width = 2.5 in, height = 2 in]{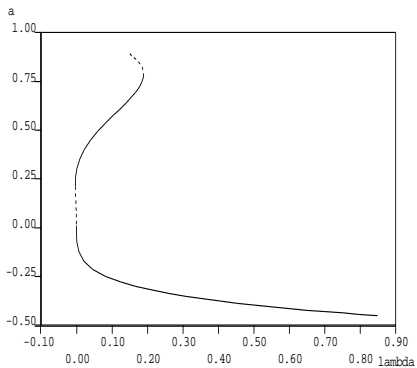}

\vspace{-.25 in}

		\caption{Continuation of the detuned system in the parameter 
$\lambda$.} 
\label{continue}
	\end{centering}
\end{figure}

\begin{figure}[htb] 
	\begin{centering}
		\leavevmode
		\includegraphics[width = 2.5 in, height = 2 in]{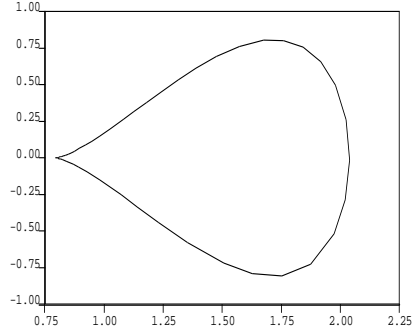}

\vspace{-.25 in}

		\caption{Right homoclinic orbit for $\lambda = 0.1887$.} 
\label{right}
	\end{centering}
\end{figure}

\begin{figure}[htb] 
	\begin{centering}
		\leavevmode
		\includegraphics[width = 2.5 in, height = 2 in]
{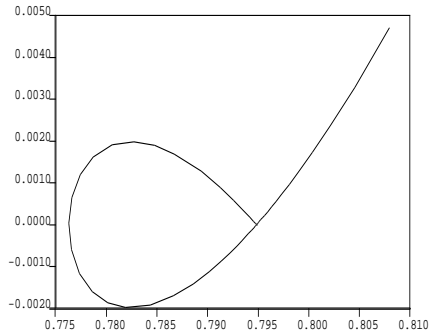}
\vspace{-.25 in}

		\caption{Left homoclinic orbit for $\lambda = 0.1887$.} 
\label{left}
	\end{centering}
\end{figure}

	In general, AUTO has difficulties near cusps.  As with DSTool, one can
 approximate the cuspidal homoclinic orbit using AUTO.  In order to do this, 
one provides initial values for the HomCont continuation parameters 
(corresponding to the initial point in the $(a,P)$-plane) to the right of the 
saddle point $(a^*,0)$.  This allows AUTO to continue along the homoclinic 
orbit for values of $\lambda$ closer to $\bar{\lambda}$ than if one had 
started as close to the saddle point as machine precision would allow.  For 
the present example, the closest accurate plot we obtained was for 
$\lambda = 0.1887$.  The right homoclinic orbit is pictured in Figure 
\ref{right} and the left one is pictured in Figure \ref{left}.  Observe that 
the one on the right \begin{itshape}looks\end{itshape} like it has a cusp at 
the saddle point and that the left one is very small.  As the saddle-center is
 approached, the left homoclinic orbit shrinks to a single point and the right
 one becomes more cusplike.

\section{Chaotic Case: Presence of Coupling}

\vspace{-.05 in}

	There are other methods one can use to perform analysis near the cusp.
  One can, for example, use shooting methods.\cite{rod}  (AUTO uses a 
collocation method.)  Saddle-centers are a degenerate case of Takens-Bogdanov 
(TB) bifurcations\cite{gucken}, so one can add dummy parameters and analyze 
the cuspidal loop by computing the locus of a TB bifurcation in parameter 
space, moving along the TB curve until one finds a Hopf bifurcation of another
 equilibrium, and then following the evolution of the periodic orbit created 
in the Hopf bifurcation as the parameters follow the TB bifurcation curve.  If
 a cuspidal loop exists, this method will find it when the periodic orbit 
collapses into the cusp point.

	For the chaotic case, the equilibrium points satisfy $x = P = y = 0$, 
$z = \pm 1$, and $\frac{\partial V}{\partial a} = \frac{2}{a^3} (\epsilon_+ 
\pm \epsilon_-)$, where the factor of $\pm 1$ in the last quantity corresponds
 to the equilibrium value of $z$.  The eigenvalues for the equilibria of the 
chaotic system (Eqs. \ref{xdot}--\ref{Pdot}) are of the form
\begin{equation}
	\sigma = 0, \text{         } \pm \frac{\sqrt{-2M (\alpha \pm 
\sqrt{\beta})}}{2M a^2},
\end{equation}
where both $\alpha$ and $\beta$ have terms whose signs depend on whether $z$ 
is $1$ or $-1$.  The quantities $\alpha_\pm$ and $\beta_\pm$ are given by
\begin{equation}
	\alpha_+ =  a^4 \frac{\partial^2 V}{\partial a^2} + \omega_0^2M - 
4\mu^2\epsilon_- + 6(\epsilon_+ + \epsilon_-),
\end{equation}
\begin{equation}
	\alpha_- = a^4 \frac{\partial^2 V}{\partial a^2} + \omega_0^2M + 
4\mu^2\epsilon_- + 6(\epsilon_+ - \epsilon_-),
\end{equation}
\begin{equation}
	\beta_+ = I_1 + I_2^+ + I_3^+ + I_4^+,
\end{equation}
where
\begin{gather}
	I_1 \equiv a^8 \left(\frac{\partial^2 V}{\partial a^2}\right)^2, 
\notag \\
	I_2^+ \equiv \frac{\partial^2 V}{\partial a^2}\left(12a^4\epsilon_+ - 
2\omega_0^2Ma^4 + 12a^4\epsilon_- - 8\mu^2 a^4 \epsilon_-\right), \notag \\ 
	I_3^+ \equiv 16\mu^4\epsilon_-^2 - 8\omega_0^2M\mu^2 - 
48\mu^2\epsilon_-^2 - 12\omega_0^2M\epsilon_+ + 72 \epsilon_+\epsilon_- 
- 12\omega_0^2M\epsilon_-, \notag  \\ 
	I_4^+ \equiv - 48 \mu^2\epsilon_+\epsilon_- + \omega_0^4M^2 + 
36(\epsilon_+^2 + \epsilon_-^2), 
\end{gather}
and
\begin{equation}
	\beta_- = I_1 + I_2^- + I_3^- + I_4^-,
\end{equation}
where $I_1$ is as before,
\begin{gather}
	I_2^- \equiv \frac{\partial^2 V}{\partial a^2}\left(12a^4\epsilon_+ - 
2\omega_0^2Ma^4 - 12a^4\epsilon_- + 8\mu^2 a^4 \epsilon_-\right),  \notag \\ 
	I_3^- \equiv 16\mu^4\epsilon_-^2 + 8\omega_0^2M\mu^2 - 
48\mu^2\epsilon_-^2 - 12\omega_0^2M\epsilon_+ - 72 \epsilon_+\epsilon_- + 
12\omega_0^2M\epsilon_-,  \notag \\ 
	I_4^- \equiv 48 \mu^2\epsilon_+\epsilon_- + \omega_0^4M^2 + 
36(\epsilon_+^2 + \epsilon_-^2).
\end{gather}

	Analogous to the planar system, only a generalization of saddle-center
 bifurcations can occur.  As the energy is increased, a bifurcation 
corresponds to an increase in the dimension of the center manifold by two as a
 pair of real eigenvalues of opposite signs becomes a pair of pure imaginary 
eigenvalues.  For a double-well potential, the only possibility is a jump in 
the dimension of the center manifold from three to five.  For more complicated
 potentials, there may be parameter values with a one-dimensional center 
manifold.  As before, one can determine the location of this bifurcation by 
finding the equilibria for which $H(a,P)$ has a double root.  One 
again finds that the equilibrium point $(0,0,\pm 1,\bar{a},0)$ at the 
bifurcation satisfies
\begin{equation}
	\frac{\partial V}{\partial a} \left(\bar{a},0 \right) = 
-\frac{\bar{a}}{3} \frac{\partial ^2 V}{\partial a^2} \left(\bar{a},0 \right).
\end{equation}

	Finding the parameter values at which this occurs is not the simple 
issue it was in the previous case.  In the planar case, $\lambda$ was a 
probabilistic weighting of two energies, so it could be varied continuously 
past the bifurcating value $\bar{\lambda}$.  However, the parameters 
$\epsilon_1$, $\epsilon_2$ in the present system are part of a discrete energy
 spectrum, and so one cannot vary them continuously.  In practice, therefore, 
this model predicts superposition states on each side of the bifurcation, but 
one does not expect to observe the system \begin{itshape}at\end{itshape} a 
bifurcating value of $(\epsilon_1,\epsilon_2)$ because the set of all 
$(\epsilon_1,\epsilon_2)$ that correspond to bifurcating values has measure 
zero.  (One could vary $V_2$ continuously if one wanted to examine 
bifurcations corresponding to a change in the quartic potential.  If $V_2$ is 
negative and sufficiently small for a given $V_4$ or if it's positive, the 
system will not exhibit a saddle-center.)  Additionally, numerical 
observations indicate that the bifurcation occurs at low energies 
(corresponding to superpsoitions of low quantum number states), so that for a 
given billiard system, most superposition states will exhibit an evolution 
with a five-dimensional center manifold.  The bifurcation under study may thus
 occur as one considers superpositions of increasingly excited states of the 
quantum billiard.

\begin{figure}[htb] 
	\begin{centering}
		\leavevmode
		\includegraphics[width = 2.5 in, height = 3 in]{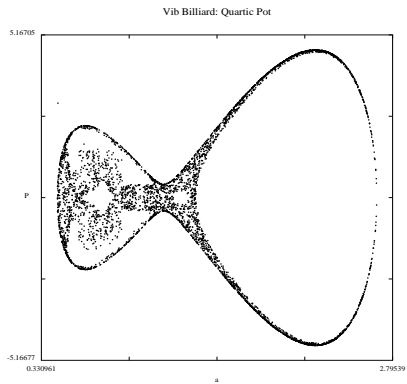}

\vspace{-.5 in}

		\caption{Poincar\'e section projected into the $(a,P)$-plane 
demonstrating that there must exist a homoclinic orbit.} \label{tubby}
	\end{centering}
\end{figure}

\begin{figure}[tb] 
	\begin{centering}
		\leavevmode
		\includegraphics[width = 2.5 in, height = 3 in]{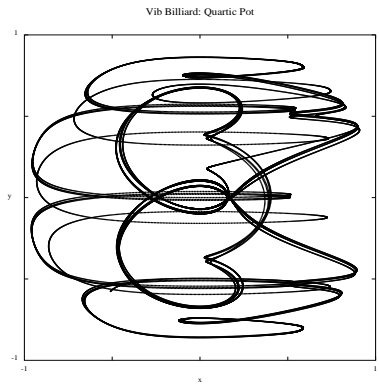}

\vspace{-.5 in}

		\caption{A homoclinic tangle projected onto the $(x,y)$-plane 
of the Bloch sphere.} \label{tangle}
	\end{centering}
\end{figure}

	Numerical investigations have shown that homoclinic orbits must exist 
for this five-dimensional system (Fig. \ref{tubby}).  Slight perturbations 
away from the homoclinic orbits lead to homoclinic tangles (Fig. 
\ref{tangle}), which are traditionally analyzed using symbolic dynamics.  The 
details have not been worked out, but the existence of a homoclinic orbit for 
the present case can be shown analytically as follows.  There is only one 
saddle point, so any saddle connection would have to be a homoclinic 
connection.  By Hamiltonian symmetry and the existence of a center to the 
right of the ``saddle-like'' (in the sense that it has one-dimensional stable 
and unstable manifolds) equilibrium, there must exist a structure to its right
 that \begin{itshape}looks\end{itshape} like a homoclinic orbit when projected
 into the $(a,P)$-plane.  It may not be a homoclinic orbit, because one 
must consider the value of $(x,y,z)$ at the point $(\tilde{a},0)$ where the 
projection intersects the $a$-axis.  One thereby considers the 2-sphere $S^2$ 
and two trajectories on it that start at the same point.  One curve begins at 
$t = \infty$ from the stable manifold, and the other starts at $t = - \infty$ 
from the unstable manifold.  One looks at the intersection of one trajectory 
at time $t$ with the other at time $-t$.  If one can prove that such a point 
exists at a time $T$, then repeating the argument shows that such an 
intersection occurs at infinitely many times.  To complete the proof, one must
 show that $P = 0$ at one of these points.

\begin{figure}[tb] 
	\begin{centering}
		\leavevmode
		\includegraphics[width = 2.5 in, height = 3 in]{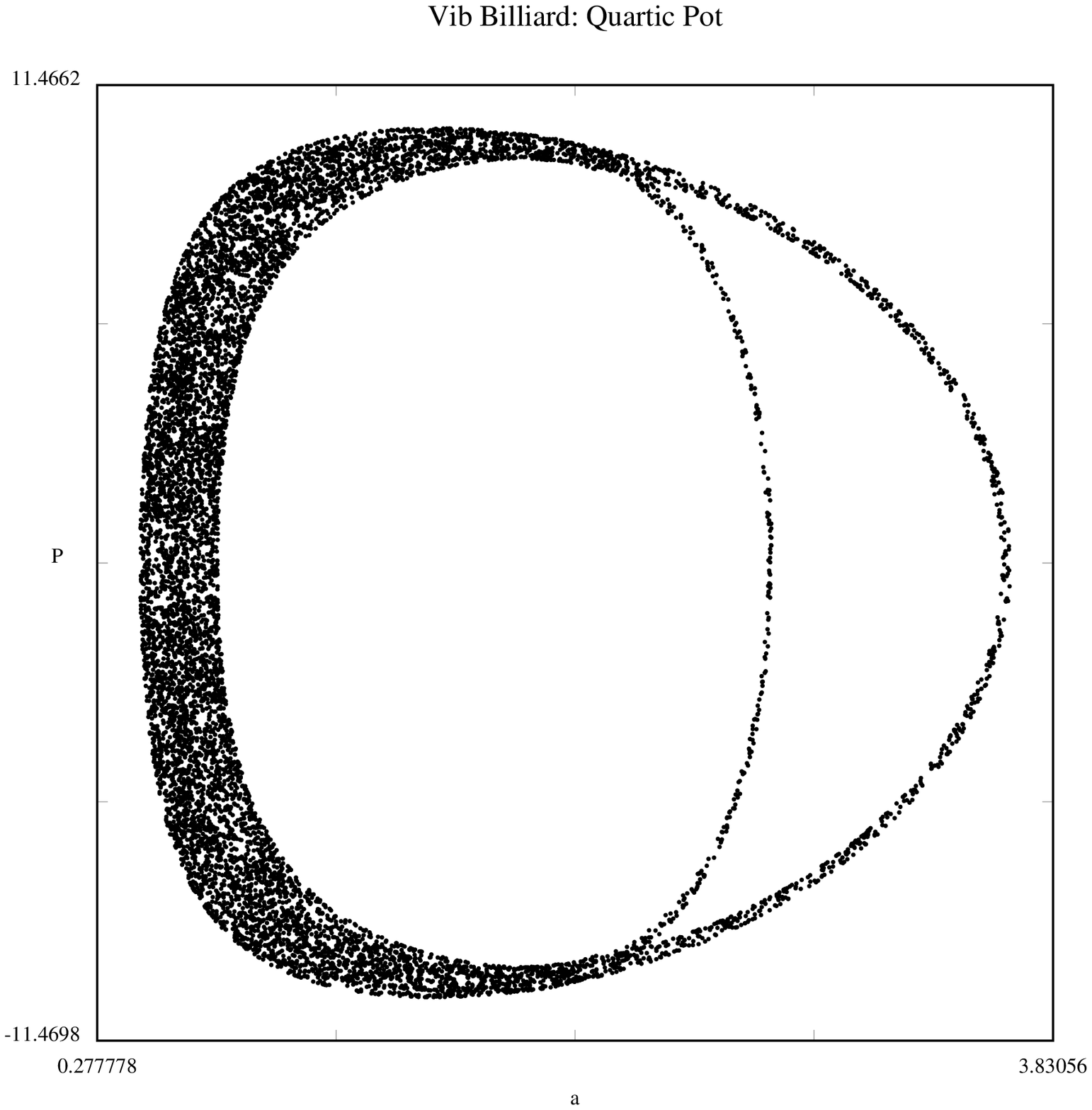}

\vspace{-.5 in}

		\caption{Chaotic Poincar\'e maps in the $(a,P)$-plane for 
billiards in both a quartic and a quadratic potential.  The plot from the 
quartic potential is the one with a smaller maximum value for the distance 
$a$.  Notice that the trajectory corresponding to the quartic potential 
generically has a larger radius of curvature.} \label{stiff}
	\end{centering}
\end{figure}

	Note that the homoclinic orbit for the chaotic system 
(Eqs. \ref{xdot}--\ref{Pdot}) emanates and terminates from a nonhyperbolic 
equilibrium, which increases the difficulty of numerical 
studies.\cite{homcont,detect}  As with the planar system discussed earlier, 
there is also a cusplike structure as the stable and unstable manifolds 
coalesce along the $a$-axis.

	As an anology, consider two undamped springs, one with a linear 
restoring force ($F_1 = -kx$) and a stiffer spring with a cubic one 
($F_3 = -kx^3$).  These two springs (with mass $m = 1$) are described, 
respectively, by the differential equations
\begin{equation}
	\ddot{x} + kx = 0,
\end{equation}
and
\begin{equation}
	\ddot{x} + kx^3 = 0. \label{cube}
\end{equation}
One observes the same stiffness phenomena for these spring systems as we did 
for the integrable case of the vibrating billiard.  (For example, the phase 
space trajectory of the cubic spring (Eq. \ref{cube}) has a larger radius of 
curvative than the analogous one in the linear spring system.  There is a 
correspondence with the other properties we discussed as well.)  By analogy 
with mass-spring systems, we thus conclude that it is reasonable to consider 
the ``stiffness'' of the potential in which a quantum billiard resides as an 
object of interest.

	The analogy with spring stiffness carries through in the chaotic case 
as well as in the planar case.  For the same initial conditions, we consider 
the potentials
\begin{equation}
	V(a) = V_2(a - a_0)^2
\end{equation}
and
\begin{equation}
	V(a) = V_4(a - a_0)^4,
\end{equation}
where for ease of comparison, $V_2 = V_4$.  As shown in Fig. \ref{stiff}, 
the quartic potential gives trajectories with a larger radius of curvature 
than those in the quadratic potential.  It thus makes sense to consider a 
potential's stiffness in the chaotic case as well as in the integrable one.

\vspace{-.05 in}

\section{Conclusions}

	We showed that saddle-centers are the only type of bifurcations that 
can occur in one \begin{itshape}dov\end{itshape} quantum billiards.  When 
there is no coupling in the superposition state, we showed how to analyze the 
resulting planar system analytically and numerically.  Considering a 
double-well potential as a representative example, we demonstrated how to 
locate bifurcations in parameter space.  We also discussed how to approximate 
the cuspidal loop using AUTO as well as how to continue past it by detuning 
the dynamical system.  We also mentioned a shooting method for a more detailed
 analysis of the cuspidal loop.  Moreover, we showed that when there is 
coupling, the resulting five-dimensional system---though chaotic---has a 
similar underlying structure.  We verified numerically that both homoclinic 
orbits and cusps occur.  Small perturbations of the system reveal homoclinic 
tangles that typify chaotic behavior.

\vspace{.1 in}

\section{Acknowledgements}

\vspace{-.1 in}

	We would like to thank Joel Ariaratnam, Richard Casey, John 
Guckenheimer, and Steven Wirkus for useful discussions concerning this 
project.  Additionally, we express our gratitude toward Alan Champneys
 for several productive suggestions regarding the numerics, Alejandro 
Rodr\'iguez-Luis for providing a preprint of his manuscript, and Paul Steen, 
whose guidance for this work as a project for ChE 753 (on which this paper is 
based) was particular valuable.

\bibliographystyle{plain}
\bibliography{ref}

\vspace{.5 in}

\subsection*{Figure Captions}

\vspace{.1 in}

Figure 1: Approximate cuspidal homoclinic orbit.  The initial 
point used was $(0.7886751,0.001)$, just above the equilibrium.

\vspace{.1 in}

Figure 2: Homoclinic orbits emanating from $(0.8916637,0)$ for 
$\lambda = 0.15$.  The label 12 refers to the right homoclinic orbit, and the
label 6 refers to the left one.

\vspace{.1 in}

Figure 3:Continuation of the detuned system in the parameter $\lambda$.

\vspace{.1 in}

Figure 4: Right homoclinic orbit for $\lambda = 0.1887$.

\vspace{.1 in}

Figure 5: Left homoclinic orbit for $\lambda = 0.1887$.

\vspace{.1 in}

Figure 6: Poincar\'e section projected into the $(a,P)$-plane 
demonstrating that there must exist a homoclinic orbit.

\vspace{.1 in}

Figure 7: A homoclinic tangle projected onto the $(x,y)$-plane 
of the Bloch sphere.

\vspace{.1 in}

Figure 8: Chaotic Poincar\'e maps in the $(a,P)$-plane for 
billiards in both a quartic and a quadratic potential.  The plot from the 
quartic potential is the one with a smaller maximum value for the distance 
$a$.  Notice that the trajectory corresponding to the quartic potential 
generically has a larger radius of curvature.

\end{document}